\documentclass{article}

\usepackage{PRIMEarxiv}

\usepackage[utf8]{inputenc} 
\usepackage[T1]{fontenc}    
\usepackage{hyperref}       
\usepackage{url}            
\usepackage{booktabs}       
\usepackage{amsfonts}       
\usepackage{nicefrac}       
\usepackage{microtype}      
\usepackage{lipsum}
\usepackage{fancyhdr}       
\usepackage{graphicx}
\usepackage{url}
\graphicspath{{media/}}     

\title{A Survey of Third-Party Library Security Research in Application Software
\thanks{\textit{\underline{Citation}}: 
\textbf{Authors. Title. Pages.... DOI:000000/11111.}} 
}

\author{
  Jia Zeng, Dan Han, Yaling Zhu, Yangzhong Wang, Fangchen Weng\\
  School of Cyberspace Security\\
  Hainan University \\
  China\\
}

\begin{document}
\maketitle

\begin{abstract}
In the current software development environment, third-party libraries play a crucial role. They provide developers with rich functionality and convenient solutions, speeding up the pace and efficiency of software development. However, with the widespread use of third-party libraries, associated security risks and potential vulnerabilities are increasingly apparent. Malicious attackers can exploit these vulnerabilities to infiltrate systems, execute unauthorized operations, or steal sensitive information, posing a severe threat to software security. Research on third-party libraries in software becomes paramount to address this growing security challenge. Numerous research findings exist regarding third-party libraries' usage, ecosystem, detection, and fortification defenses. Understanding the usage and ecosystem of third-party libraries helps developers comprehend the potential risks they bring and select trustworthy libraries. Third-party library detection tools aid developers in automatically discovering third-party libraries in software, facilitating their management. In addition to detection, fortification defenses are also indispensable. This article profoundly investigates and analyzes this literature, summarizing current research achievements and future development directions. It aims to provide practical and valuable insights for developers and researchers, jointly promoting the healthy development of software ecosystems and better-protecting software from security threats.
\end{abstract}


\section{Introduction}
In the digital era, the software industry has brought many conveniences closely related to our lives. With the continuous development of technology and the exchange of community ideas, open-source software is gradually emerging. Compared to software developed by individuals or organizations in the past, open-source software adopts a collaborative development approach, where many developers share the source code of their libraries in open-source communities or package managers, allowing other users to download, use, and modify them. Other developers can then significantly improve software development efficiency and reduce production costs by installing these libraries, rapidly bringing software to the market. This scalable and flexible software development model ensures that anyone with libraries and components can modify, extend, and redistribute them, enhancing reusability and accessibility. Libraries developers share in open-source communities are typically referred to as third-party libraries (TPL), open-source software, components, or dependencies.

By sharing third-party libraries, developers can leverage the experience and knowledge of others, avoiding the need to write complex functionality or algorithms from scratch. Third-party libraries cover various areas of functionality, such as data processing, image processing, network requests, machine learning, database connections, and more. The installation and use of these libraries are typically done through package management tools (such as PyPI, npm, Maven, etc.), which make library installation and version control simple and convenient. Synopsys' "Open Source Security and Risk Analysis" report for 2023\cite{synopsys2023} emphasizes that open source is the foundation of the vast majority of software construction today. In this year's audit, the average number of open-source components increased by 13\%. This demonstrates the importance of third-party libraries in modern software development. The open-source nature of third-party libraries promotes the prosperity and collaboration of development communities, driving rapid technological advancement. At the same time, it also provides developers with more choices and flexibility, accelerating the pace of software innovation. We can build more robust, feature-rich, and efficient software applications by sharing and using third-party libraries.

In addition to convenience and flexibility, with the widespread use of third-party libraries, paying attention to the risks and compliance of third-party libraries in application software has become crucial. Ladisa et al.\cite{ladisa2023sok} analyzed the security risks posed by open-source components and categorized attacks on open-source software supply chains. The research shows that the security impact caused by open-source software is significant because the complexity of the open-source supply chain leads to a significant attack surface, providing attackers with countless opportunities to inject malicious code into open-source software, which is then downloaded and executed by victims, affecting many downstream projects. Sonatype's "2021 Software Supply Chain Report"\cite{sonatype2021} demonstrates a significant increase in attacks on open-source software supply chains, with attackers actively injecting malicious code into open-source software to shift attacks "upstream" and affect more "downstream" software. According to Google's assessment\cite{google2022}, the Log4Shell vulnerability disclosed in 2021 affected 4\% of libraries on Maven. To address the security risks brought by third-party libraries, researchers have introduced the concept of Software Composition Analysis (SCA), which is a technology and method used to identify and analyze the software components in applications. SCA aims to provide a comprehensive understanding of the software composition, including the sources, versions, licenses, security vulnerabilities, and other relevant information of third-party libraries. Third-party library detection technology is one method within SCA. It helps developers and organizations identify and manage the third-party libraries used in their applications by automating the scanning and analysis of the application's code and dependencies, assessing their potential security risks and compliance issues to enhance the security and quality of software. By comprehensively leveraging third-party library detection technology, we can ensure the reliability and security of software while rapidly innovating, providing users with a better software experience.

This article extensively collects relevant research in the field of third-party library security in application software. The selection criteria for literature are those that quantitatively and comprehensively study the use, updates, and risks of third-party libraries in application software, as well as those that propose new methods or theories for detecting third-party libraries or reinforcing defense techniques for third-party libraries in application software. Our literature collection scope includes search engines and databases such as Google Scholar, SCI (Web of Science), IEEE Explore, Springer, Elsevier, arXiv, and CNKI. We systematically read and filter the collected literature based on keywords, titles, abstracts, and other information, and further supplemented our findings by consulting references. In the end, we identified 50 relevant articles. Figure 1 shows the distribution of literature related to third-party library security research in application software. It can be observed that there is a general upward trend, especially in 2020 and 2021, with the highest number of related literature. This trend may reflect the increasing importance and widespread application of third-party libraries in software development. While the use of third-party libraries can improve development efficiency, feature richness, and code reusability, it also brings some risks and challenges, such as security vulnerabilities, dependency management, and version compatibility. The closest work to ours is the review of third-party libraries in Android application software by Xian et al.\cite{zhan2021research}, but their work only considers detection techniques for third-party libraries in Android application software and does not analyze the use of third-party libraries in software, the third-party library ecosystem, detection techniques for C/C++ software and iOS application software, and some of the latest technologies.

\begin{figure}[t]
\centering
\includegraphics[width=0.65\textwidth]{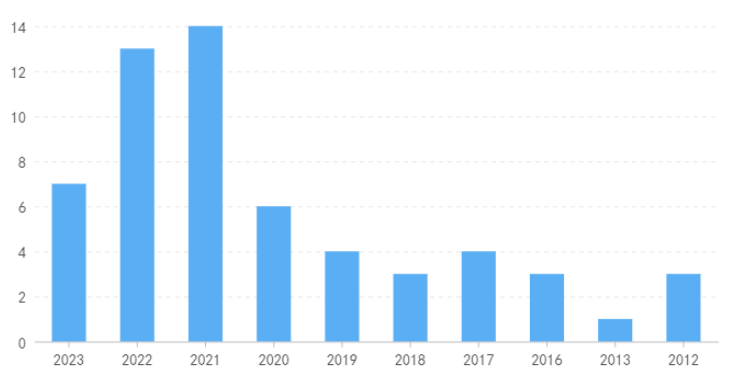}
\caption{Distribution of literature on third-party library security research in application software}
\label{fig1}
\end{figure}

Therefore, this paper provides a thorough and scientific compilation, summary, and analysis of recent research on security issues related to third-party libraries in application software. It primarily analyzes three aspects: the use, updates, and risks of third-party libraries; detection techniques for third-party libraries in software; and risk defenses for third-party libraries in software. The aim of this paper is to provide researchers and developers with an in-depth understanding of detection techniques and risks associated with third-party libraries in application software. We hope that these research findings will promote further development and innovation in this field, thereby enhancing the level of security and reliability assurance for third-party libraries in the software development process.

Section 2 of this paper provides an overview of the overall research framework. Section 3 summarizes and categorizes empirical analyses related to the use, updates, and risks of third-party libraries. Section 4 systematically analyzes detection techniques for third-party libraries in application software. Section 5 provides an analysis and summary of reinforcement defenses for third-party libraries in application software. Section 6 outlines the existing issues in this field and prospects potential research directions worthy of attention in the future.

\section{Framework for Security Research on Third-Party Libraries in Application Software}

\begin{figure}[t]
\centering
\includegraphics[width=0.55\textwidth]{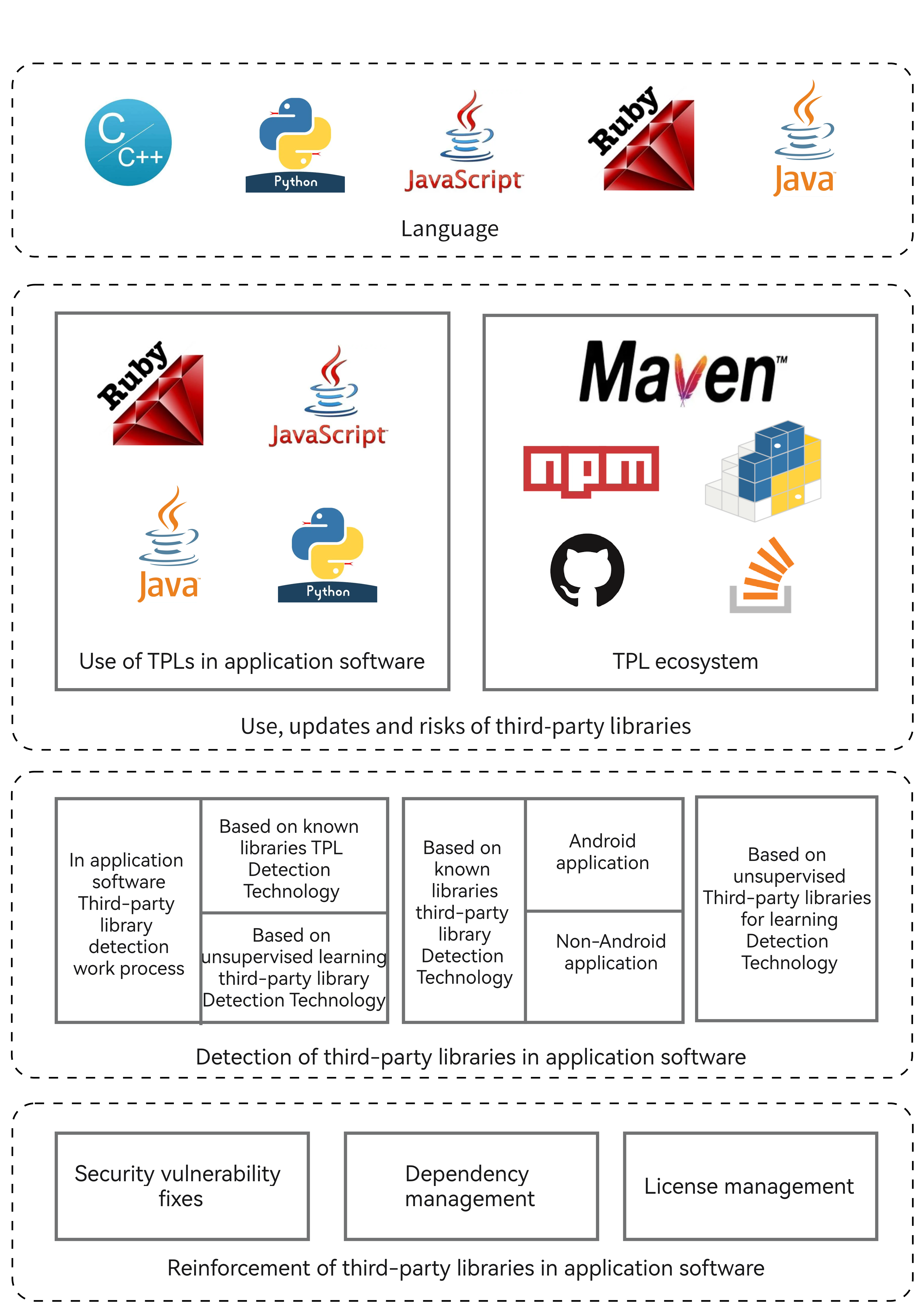}
\caption{Distribution of literature on third-party library security research in application software}
\label{fig2}
\end{figure}

The overall research framework of this paper is illustrated in Figure 2. Firstly, a quantitative and comprehensive study is conducted on the usage, updates, and risks of third-party libraries in software. By collecting relevant literature and studying practical cases, we gain insights into the usage, popularity, dependencies, and vulnerabilities of third-party libraries in software. The analysis includes the prevalence of vulnerabilities in third-party libraries, usage trends, frequency of library version updates, and developers' preferences for different libraries, providing empirical evidence for subsequent research. Secondly, this paper extensively investigates detection techniques for third-party libraries in application software. Various detection methods and technologies are explored, including those based on known libraries and unsupervised learning. The advantages and limitations of these methods are evaluated, and their feasibility and effectiveness in practical applications are discussed. Finally, this paper delves into reinforcement defenses for third-party libraries in application software. Strategies and measures for mitigating potential risks posed by third-party libraries, such as security vulnerabilities, poor code quality, and dependency management issues, are discussed. Possible solutions and recommendations are proposed.

Firstly, we will focus on the usage, updates, and risks of third-party libraries in software. This paper aims to analyze from two perspectives: firstly, from the perspective of the usage of third-party libraries in application software\cite{wang2020empirical,prana2021out,gkortzis2021software,zajdel2022open,cao2022towards}, where the asynchronous development between third-party libraries and client projects may lead to the use of outdated third-party libraries in client projects, without developers being aware of the potential risks (such as security vulnerabilities)\cite{heartbleed}. This paper will primarily analyze the intensity of usage, obsolescence, and update frequency of existing third-party libraries in software. Secondly, from the perspective of the third-party library ecosystem\cite{bommarito2019empirical,ruohonen2021large,liu2022demystifying,zerouali2022impact,mir2023effect,wu2023understanding,tang2022towards,zhang2021study}, where the third-party library ecosystem plays a crucial role in development, providing developers with rich code libraries and functionalities, greatly accelerating the development process and improving efficiency. This paper will focus on introducing some popular development languages, such as Python, Java, JavaScript, and C/C++, and existing empirical research literature on their respective third-party library ecosystems.

Next, we will focus on the detection techniques for third-party libraries in software. Current detection techniques for third-party libraries can be divided into two categories based on whether prior knowledge of the third-party libraries is required: 1) Detection techniques based on known libraries; 2) Detection techniques based on unsupervised learning. Detection techniques based on known libraries rely on pre-built databases of third-party libraries for detection. These databases contain information and features of known third-party libraries, such as library names, version numbers, file structures, source code, function call graphs, API call sequences, keywords, and dependencies. By matching the target application software with the pre-built database of third-party libraries, the third-party libraries used in the application software can be identified. The advantage of this method is its high accuracy because it relies on known information about third-party libraries. However, it also has some limitations. Firstly, maintaining an up-to-date database of third-party libraries is required to track new library versions and changes. Secondly, if the code uses closed-source or unknown third-party libraries, matching items may not be found in the known library database. Some studies use whitelists based on package names of known libraries, which are incomplete and unable to handle obfuscation. Liu et al.\cite{liu2022demystifying}, for example, extracted the names of third-party libraries and clustered them based on frequency. They then refined and collected a whitelist of third-party libraries from a large number of Android applications. On the other hand, detection techniques based on unsupervised learning do not rely on prior knowledge of third-party libraries but instead analyze the characteristics and patterns of the code itself for detection. This method utilizes machine learning algorithms to cluster, classify, or recognize patterns in the code, extracting features that may be indicative of third-party libraries. Therefore, even without prior knowledge of these libraries, third-party libraries used in the code can be identified. Detection techniques based on unsupervised learning offer flexibility and can discover closed-source, new, or unknown third-party libraries, but their accuracy may be lower, especially when dealing with complex code. Additionally, this method may be influenced by challenges such as code changes, insufficient sample data, or difficulties in feature extraction. It is important to note that these two classifications are not mutually exclusive and can be combined to improve the accuracy and coverage of third-party library detection. Different methods and techniques may be suitable for different application scenarios and requirements.

Finally, we focus on reinforcement defenses for third-party libraries in application software, with researchers striving to address the risks associated with third-party libraries in software. Existing research on reinforcement defenses for third-party libraries in software primarily revolves around three aspects: third-party library vulnerability patching, dependency management, and license management.

\subsection{The usage, updates, and risks of third-party libraries}

\subsubsection{The Use of Third-Party Libraries in Application Software}
Software developers often integrate third-party libraries into their projects to speed up development. However, the intensity of usage, obsolescence, and update frequency of third-party libraries in a project can impact its security. Several serious incidents in the real world have been triggered by third-party libraries. For instance, the Heartbleed vulnerability disclosed in 2014\cite{heartbleed}, which affected the OpenSSL library, was a significant security flaw. OpenSSL is an open-source software library widely used for encrypted communication, with many websites and web applications relying on it to secure the transmission of sensitive data. It was estimated that this vulnerability could have affected hundreds of thousands, if not millions, of websites and network services. Many well-known websites, social media platforms, email services, online stores, and other online services may have been affected by this vulnerability. Another example is the event-stream incident in 2018, where a maintainer of the event-stream package maliciously added a dependency containing malicious code to event-stream. This allowed attackers to remotely execute malicious operations through the malicious code. The primary target of this malicious code was users of the "copay" cryptocurrency wallet. When copay users used the affected event-stream package, attackers were able to obtain the private keys of copay wallets through the malicious code, potentially enabling them to steal users' cryptocurrency. This vulnerability affected many users of copay wallets.

These security incidents triggered by third-party libraries should be of great concern to developers. Developers need to review the source code of third-party libraries to ensure their quality and security. It's essential to regularly monitor and update third-party libraries to promptly apply new versions that have patched vulnerabilities. Additionally, developers should enhance their awareness of security issues, conduct security testing, and perform code audits to detect potential security vulnerabilities early on. Only by strengthening the management and maintenance of third-party libraries can the potential risks they pose be mitigated, ensuring the overall security of the software system. Therefore, in-depth research and analysis of the usage of third-party libraries in software can help improve software security.

Wang et al.'s\cite{wang2020empirical} study conducted a detailed analysis of Java open-source projects and third-party libraries. The research found that most third-party libraries only have a small portion of APIs used in projects, and multiple versions of the same library might be used in a project, leading to dependency conflicts and increased maintenance costs. Additionally, many projects adopt outdated third-party libraries, and there is a slow response to updating library versions, which increases the security risks faced by projects.

Prana et al.'s\cite{prana2021out} research focused on detecting and analyzing vulnerabilities in third-party libraries used in Java, Python, and Ruby software projects. The study found that the activity level of a project, its popularity, and the experience of developers have no correlation with handling vulnerabilities in third-party libraries. Moreover, vulnerabilities in third-party libraries in projects take a long time to be fixed, significantly impacting the security of the software.

Gkortzis et al.'s\cite{gkortzis2021software} analysis focused on the relationship between software security and code reuse. Researchers evaluated potential vulnerabilities in software projects through static analysis and found that there is no direct correlation between the number of potential vulnerabilities and the rate of code reuse. However, the number of disclosed and potential vulnerabilities is closely related to the number of dependencies included in the project. This underscores the importance of managing and maintaining dependencies, especially when using open-source libraries in software development.

Zajdel et al.'s\cite{zajdel2022open} work described the impact of using open-source software on the security of applications from the perspectives of the open-source software supply chain and developers' use of open-source software. Projects in open-source software communities may lack testing and maintenance, leading to numerous security issues. To mitigate the risks associated with open-source software, the authors proposed using Software Bill of Materials (SBOM) to track the introduction and usage of open-source software and verified the feasibility of SBOM.

Cao et al.'s\cite{cao2022towards} study identified three dependency issues—missing dependencies, redundant dependencies, and inconsistent version constraints—that could have serious consequences. They proposed a tool called PyCD to accurately extract dependency information from configuration files and conducted an in-depth empirical study on Python projects. The study discussed the prevalence, causes, and evolution of these dependency issues, reported harmful instances of dependency issues to developers, and received feedback, demonstrating the negative impact of these dependency issues on project maintenance. The results suggest that developers should pay attention to these dependency issues.

The aforementioned studies analyze the importance of using third-party libraries in software development and the potential security risks associated with their usage. While third-party libraries can expedite development, excessive usage, outdated library versions, and lack of timely updates can lead to serious security incidents. Several past security vulnerability cases have underscored the criticality of managing and maintaining third-party libraries. Research indicates that many projects utilize only a small portion of the APIs provided by third-party libraries and face issues with dependency conflicts, which developers should address. Additionally, projects that lag in responding to third-party library vulnerability fixes and lack code reuse also elevate security risks. The comprehensive findings of these studies suggest that developers should enhance scrutiny, regularly monitor and update third-party libraries, raise security awareness, conduct security testing, and perform code audits. Furthermore, introducing Software Bill of Materials (SBOM) to track the usage of open-source software can help mitigate risks associated with such software. Through in-depth research and analysis of third-party library usage, overall software system security can be effectively enhanced.

\subsubsection{Third-Party Library Ecosystem}

The third-party library ecosystems of various programming languages have had a significant impact on software development. Empirical research helps understand and evaluate the characteristics, trends, and potential issues of these ecosystems, providing better resources and guidance for developers and researchers. Python has a vast and active third-party library ecosystem, with PyPI (Python Package Index) being the most well-known. Existing empirical studies mainly focus on PyPI's usage patterns, quality assessments, and version management of different libraries. In the JavaScript language, the most popular third-party library ecosystem is npm (Node Package Manager). npm, as one of the essential tools in the JavaScript community, allows developers to install, publish, and manage JavaScript packages. Existing research literature focuses on trends in library usage, quality assessments, and vulnerability propagation. In the Java language, the most popular third-party library ecosystem is Maven. Maven, as a building and dependency management tool, provides convenient ways for Java developers to introduce and manage libraries. Existing empirical studies mainly focus on library classification, version management, vulnerability propagation, and usage patterns. The third-party library ecosystem in C/C++ is quite diverse, with no clearly dominant ecosystem but many options such as Conan, CMake, etc. Most C/C++ third-party libraries are still stored on platforms like GitHub, so existing literature on C/C++ ecosystem analysis mainly focuses on the quantity, activity, usage patterns, and quality assessments of third-party libraries in open-source communities.

In-depth research on the PyPI ecosystem conducted by Bommarito et al. provided a comprehensive analysis. The study found that PyPI is experiencing accelerated growth in both size and rate. Distribution of packages, versions, authors, and import statements exhibited a highly right-skewed pattern, indicating that a minority of objects represent the majority. Most packages are in immature or unstable states, primarily serving developers or researchers. Over half of the packages adopt permissive licenses such as MIT, BSD, or Apache, with only a few using GPL licenses. Moreover, more than 25\% of packages lack clear license metadata, potentially posing risks to users. Additionally, a complex dependency network has formed within PyPI, with the number of import statements growing by 61.7\% annually, with most referring to standard libraries or duplicate objects. These research findings offer valuable insights into understanding the development trends, license usage, and dependency management of Python packages. In addition to analyzing data on packages, versions, authors, licenses, and dependencies, researchers also examined security issues within the PyPI ecosystem. Ruohonen et al. used the static analysis tool Bandit to detect security issues in Python packages on PyPI. Their study revealed that 46\% of PyPI packages have at least one security issue. The most common security issue types include hard-coded passwords, improper exception handling, and code injections (such as command injection, SQL injection, etc.). Importantly, the study also found a positive correlation between package size and the number of security issues, indicating that larger packages tend to have more security problems. This research underscores the importance of addressing software security within the ecosystem. PyPI, as a core component of the Python ecosystem, directly impacts the security of numerous Python developers and applications. Through the use of static analysis tools, this study identified various security issues within PyPI packages, highlighting some of the most common and dangerous security issue types. The research findings provide valuable insights and guidance for improving the software security of the PyPI ecosystem.

In their research on the npm ecosystem, Liu et al.\cite{liu2022demystifying} investigated the propagation and evolution of vulnerabilities in the npm dependency tree. They proposed a knowledge graph-based dependency resolution method capable of statically and accurately parsing dependency trees at any given time point, along with their vulnerability propagation paths. Based on this method, they conducted a large-scale empirical study of the npm ecosystem, revealing some valuable findings. For example, they found that there is a large number of vulnerabilities present in npm packages, and their impact is increasing over time. Most vulnerabilities are introduced into the dependency tree before they are discovered, and the majority of fixed versions are released before they are disclosed. However, there are still some vulnerabilities that cannot be completely eliminated from the dependency tree, and some user projects may inevitably include vulnerabilities. They also proposed several solutions tailored to different stakeholders, such as vulnerability-fixing methods based on the dependency tree, which can eliminate more vulnerabilities than official tools. This literature holds significant importance and insights for enhancing the security of the npm ecosystem. Zerouali et al.\cite{zerouali2022impact} conducted an analysis of security vulnerabilities in two ecosystems, npm and RubyGems, comparing their quantities, severity, types, disclosure times, fix statuses, and impacts on dependents. They also suggested some recommendations for improving package security and future research directions. The study found that the number of vulnerabilities in npm grows exponentially, while in RubyGems, it grows linearly. Moreover, the severity of vulnerabilities in npm is higher than in RubyGems. Most affected package versions have not been fixed or require updating to incompatible versions for fixes. This implies that customers relying on these packages should use the latest available versions, especially npm customers. The research suggests that software developers should promptly update their dependencies and use security tools to detect potential risks. Future research directions could explore the exploitability of vulnerabilities, propagation paths, and scope of impact.

In their in-depth research on the Maven ecosystem, Mir et al.\cite{mir2023effect} conducted a thorough investigation into the propagation of library vulnerabilities within the Maven ecosystem. By considering the transitivity of dependency relationships and granularity, they evaluated the impact of vulnerabilities on projects. The experimental results showed that transitivity significantly affects the propagation of vulnerabilities, with most vulnerabilities spreading to projects through transitive dependencies. Granularity is also crucial, as less than 1\% of packages directly invoke vulnerable code in their dependencies. In popular Maven projects, vulnerabilities may pose higher security risks to other dependent projects. Limiting the depth of dependency tree resolution is an effective technique for reducing the computational time required for vulnerability analysis. Wu et al.\cite{wu2023understanding} studied the threat posed by upstream vulnerabilities in the Maven ecosystem to downstream projects. Their research analyzed various aspects of upstream vulnerabilities, including their accessibility, exploitability, responses from downstream projects, and awareness among developers. The results indicated that the majority of upstream vulnerabilities do not actually affect downstream projects. However, certain commonly used APIs may threaten a large number of downstream projects. The exploitability of upstream vulnerabilities is influenced by factors such as path length, path constraints, and calling context. Additionally, downstream developers exhibit varying responses and evaluation methods towards upstream vulnerabilities.

In their in-depth research on the C/C++ third-party library ecosystem, Tang et al.\cite{tang2022towards} investigated dependencies within the C/C++ ecosystem. The study outlined the lack of a unified package manager and dependency detection tools for C/C++ and proposed an analysis framework based on dependency lifecycle, which includes third-party library data sources, reuse methods, and toolchains. Additionally, they designed and implemented a comprehensive C/C++ dependency scanner called CCScanner, capable of handling 21 package management tools and code clone methods. Through the use of CCScanner, they conducted a large-scale empirical study on 24,000 GitHub C/C++ repositories, exploring characteristics, data scope, critical libraries, and version constraints of dependencies within the C/C++ ecosystem, providing insights and recommendations for practitioners and future research.

The research findings indicated several issues within the C/C++ ecosystem, including confusion in dependency management, data fragmentation, the importance of system libraries, and the influence of version selection, all of which require further attention and improvement. Zhang et al.\cite{zhang2021study} focused on security vulnerabilities in C/C++ code snippets on Stack Overflow. They used the Cppcheck tool to scan these code snippets and identified 32 different types of security vulnerabilities, accounting for 36\% of all security vulnerability types in C/C++. The study also revealed that some security vulnerabilities in these code snippets were associated with actual software system vulnerabilities, while others had not been reported as vulnerabilities. The proportion of code snippets containing security vulnerabilities doubled from 2008 to 2018. Only 7.5\% of users contributed code snippets with security vulnerabilities, and more active users contributed fewer security vulnerabilities. Some suggestions from the Stack Overflow community were adopted. The study also proposed recommendations for improving the quality of Stack Overflow as an information source and code-sharing platform. The third-party library ecosystems in various programming languages have had a significant impact on software development. Empirical research literature provides valuable insights for understanding and evaluating the characteristics, trends, and potential issues of these ecosystems. Research on the PyPI ecosystem has revealed the trend of its growing scale and characteristics such as packages, versions, and authors. Additionally, the analysis of security issues emphasizes the importance of focusing on software security within the ecosystem. In empirical studies of the JavaScript ecosystem, attention has been given to the propagation and evolution trends of vulnerabilities in npm packages. Similarly, research on the Maven ecosystem has focused on the impact of library vulnerability propagation and techniques to limit parsing of dependency trees. In the case of the C/C++ third-party library ecosystem, research has delved into the characteristics of dependencies, data scope, and issues related to reuse. These research findings provide important guidance and recommendations for enhancing the security and quality of third-party library ecosystems.

\section{Third-party library detection in application software}
\subsection{Workflow for third-party library detection in application software}
(1) Workflow for third-party library detection based on known libraries

Different types of third-party library detection techniques have their own unique workflows, but similar workflows may be observed across different applications for the same type of technique. In existing third-party library detection techniques based on known libraries, the common workflows can be divided into three key modules: data collection module, data processing module, and library matching module.

During the data collection phase, third-party library data is first collected from multiple sources, such as public software package registries, version repositories, and community-maintained library lists. The third-party library data collection module needs to gather necessary information about the third-party libraries, including source code for each version, configuration files, build scripts, and related dependency files. The objective of this module is to obtain relevant data about the third-party libraries used in the software, and the collected information will be utilized in subsequent analysis and detection processes.

In the data processing phase, there are mainly two parts. First, the third-party library data collected by the data collection module is cleaned, deduplicated, and standardized to ensure the accuracy and consistency of the data. Then, feature extraction is performed on this data. Most studies extract syntax and semantic information from third-party libraries as code features. Finally, the extracted third-party library features are used to construct a third-party library database for subsequent detection processes. Additionally, third-party library files may depend on other third-party libraries, known as nested third-party libraries. However, nested third-party libraries may lead to false positives. Some research methods segment the code of third-party libraries after feature extraction to remove nested third-party libraries. Second, data processing is carried out for the target software. Existing research can detect third-party libraries in Android applications, C/C++ software, and iOS applications. Typically, these types of target software are processed similarly to third-party library data processing. However, for Android applications, decompilation is required before feature extraction. Some research methods also decouple the target software modules before extracting features to build candidate third-party libraries for higher accuracy. Literature\cite{zhan2021research} mentions that the purpose of module decoupling is to split non-core modules of the application into different candidate third-party libraries, where non-core modules refer to those composed of third-party libraries within the application. In other words, it aims to determine a unique third-party library boundary that does not use other third-party libraries.

In the library matching phase, known-library-based third-party library detection techniques match the features extracted from the target software with the third-party library features in the third-party library database to derive a list of third-party libraries used by the target software. Additionally, some tools not only detect third-party libraries but also identify the versions of third-party libraries used in the software. These tools typically use weighted allocation methods to determine the versions of third-party libraries used.

In summary, known-library-based third-party library detection techniques mainly focus on improving feature selection methods in the data processing phase and similarity comparison methods in the library matching phase to enhance the accuracy and efficiency of third-party library detection.

In the data processing phase, the goal of feature selection is to extract the most representative information from both the target software and the third-party libraries, and transform it into features for subsequent similarity recognition between the target software and the third-party libraries by the library identification module. Specifically, feature selection involves two aspects of information: syntax and semantics. Regarding syntax information, the following features can be extracted: third-party library name, class name, function name, character name, variable name, function centroid, package structure, class signature, function signature, and basic blocks of control flow graphs. These features mainly focus on the structure and naming conventions of the code and can be directly extracted from the third-party library data. Regarding semantic information, the following features can be extracted: control flow graph, API call graph, class dependency graph, etc. These features describe the semantic relationships and interactions between the code and require further analysis and interpretation of the third-party library data. However, because some software may undergo obfuscation, obfuscation tools can change syntax information such as third-party library names, class names, function names, character names, and variable names. Depending solely on syntax information may lead to false negatives. Reference\cite{wang2020empirical} uses more semantic information in feature extraction, including control flow graphs and class dependencies, to improve robustness against code obfuscation. Reference\cite{zhan2021research} found that by including more semantic code features and utilizing class dependencies to build TPL instances, better robustness against code obfuscation can be achieved. Since obfuscation tools may alter semantic relationships such as package structure, control flow graphs, and interface classes, most studies adopt a comprehensive approach to extract both syntax and semantic information. This approach can more comprehensively describe the features of third-party libraries and enhance robustness against code obfuscation. Therefore, in the data processing phase, the goal of feature selection is to extract syntax and semantic information from the data and combine them to form the features of third-party libraries. Such a comprehensive approach can improve the tool's resistance to obfuscated code, thereby enhancing the accuracy and efficiency of third-party library detection.

In the library matching phase, it's essential to choose the appropriate matching granularity and method that are suitable for the current problem. Common matching granularities include code-level, basic block-level, method-level, class-level, package-level, etc. Most research methods combine multiple granularities for similarity comparison. There are also various methods available for matching comparison. Commonly used methods include weighted feature matching algorithms, fuzzy class matching, hierarchical indexing, LSH (Locality Sensitive Hashing), ssdeep, and so on. During the library matching phase, it's crucial to select the appropriate similarity comparison granularity and method based on the specific problem at hand, in order to enhance the accuracy and efficiency of third-party library detection. Different granularities and methods can capture different levels of code similarity and comprehensively consider code structure, semantic relationships, and functional features, thus enabling a more comprehensive library matching process.

(2) The workflow of third-party library detection technology based on unsupervised learning

The workflow of third-party library detection technology based on unsupervised learning is different from that based on known libraries. As shown in Figure 3. It utilizes clustering methods to detect third-party libraries in software. In contrast to techniques based on known libraries, the unsupervised learning approach does not rely on known third-party library information. Instead, it automatically discovers existing third-party libraries by conducting clustering analysis on software data. Below, we will elaborate on the workflow of third-party library detection technology based on unsupervised learning.

\begin{figure}[t]  
  \centering
  \includegraphics[width=0.9\textwidth]{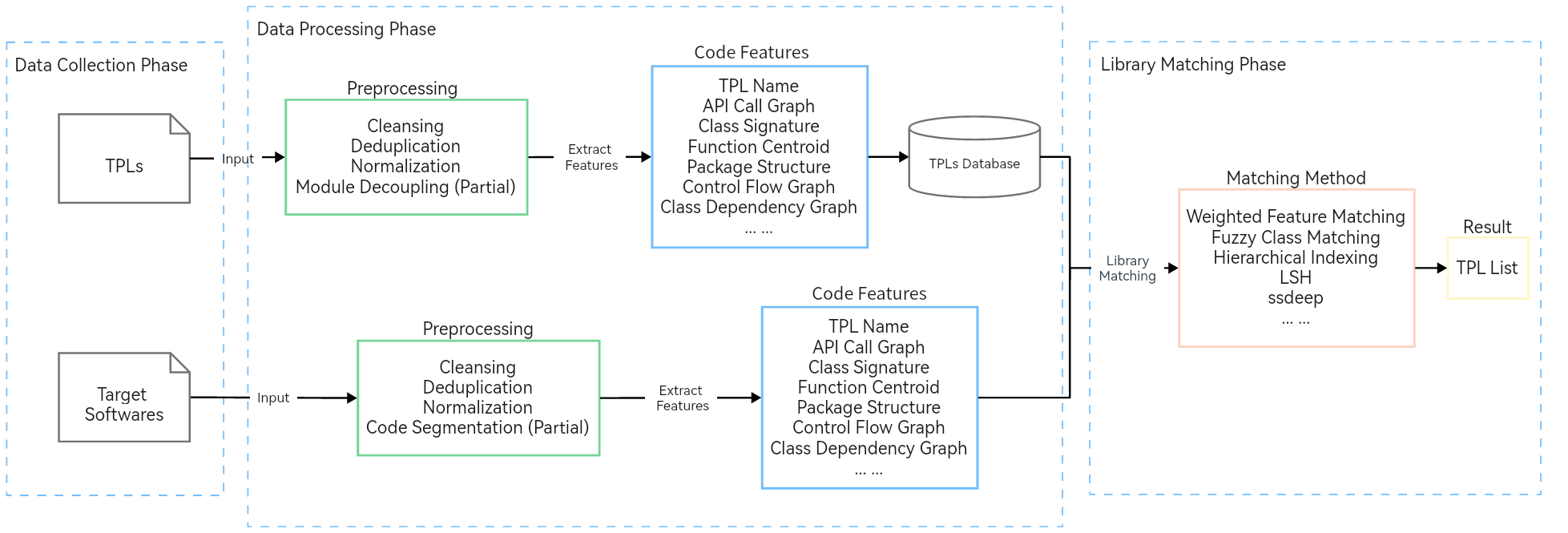}
  \caption{Workflow of TPL detection technology based on known database.}
  \label{fig3}
\end{figure}

In the third-party library detection technology based on unsupervised learning, the first step is to collect software data. This involves obtaining the source code or binary files of the target software for subsequent analysis and processing. Next, feature extraction is performed on the collected software data. The goal of feature extraction is to extract representative features from the software data for subsequent clustering analysis. Common features include code structure, naming conventions, function call relationships, etc. These features can reveal the presence and characteristics of third-party libraries in the software, providing a basis for clustering analysis. Subsequently, clustering analysis methods are used to cluster the extracted features. The objective of clustering is to group software parts with similar features into clusters, thereby discovering potential third-party libraries. Clustering methods can be distance-based, such as K-means clustering, or density-based, such as DBSCAN. Finally, the software parts classified as third-party libraries are determined based on the clustering results. Each category in the clustering results can be considered as a potential third-party library. Further analysis and validation can be conducted to confirm whether these categories truly represent third-party libraries, and additional processing or identification can be performed accordingly.

In conclusion, unsupervised learning-based third-party library detection technology discovers potential third-party libraries by clustering the features in software data. Its workflow includes software data collection, feature extraction, and clustering analysis. Through this unsupervised learning approach, unknown third-party libraries can be detected, providing valuable information for software analysis and security assessment.

Based on the analysis above and recent related research, it can be concluded that existing third-party library detection techniques mostly employ various feature extraction methods and integrate syntax and semantic information. These features are used to compare the similarity between third-party libraries and target software. The selected features in existing research methods are cross-cutting, making it difficult to classify existing research from the perspective of third-party library feature extraction. Therefore, it is believed in this paper that classifying third-party detection techniques based on whether prior knowledge of third-party libraries is required and the type of target software is a reasonable and meaningful approach. Such classification methods contribute to a better understanding and application of different categories of third-party library detection techniques.

\subsection{Detection Technology of Third-Party Libraries in Application Software}
In this section, we will provide a detailed overview of third-party library detection technology. Firstly, we will categorize it broadly based on whether prior knowledge of third-party libraries is required. We will begin by introducing the detection techniques based on known libraries, followed by those based on unsupervised learning. Subsequently, we will further categorize them based on different types of target software. We will start by discussing third-party library detection techniques for Android applications and then move on to those for non-Android applications. Finally, we will visually summarize all the mentioned research works, illustrating the relationships among these technologies. This visualization will aid readers in better understanding the research directions and methodologies of third-party library detection technology.

(1) Third-party library detection technology based on known libraries

This section primarily summarizes and analyzes existing third-party library detection techniques for Android applications and non-Android applications.

1) Third-party Library Detection Techniques for Android Applications

In the realm of third-party library detection techniques for Android applications, researchers have proposed various methods and tools. Some studies primarily focus on the module decoupling aspect of data processing, using module decoupling to deconstruct candidate third-party libraries, and employing simple similarity comparison methods in the library matching phase. Module decoupling is aimed at addressing nested third-party libraries, with the main objective of dividing the application into main and non-main modules. The main module defines the core functionality of the application, while the non-main modules are contributed by third-party libraries. Initially, Android applications are decoupled into separate modules, where the code within each module (main or non-main) is highly coupled, while the code between modules is loosely coupled or even independent. This is based on the premise that libraries are typically coded in a modular manner within the software context, resulting in low coupling but high cohesion within the library code. Coupling refers to the interdependence between modules, while cohesion describes the degree of relatedness between functions within a single module. Low cohesion implies that a given module performs tasks that are less related to each other, potentially causing issues as the module grows larger. The third-party libraries used in Android applications follow the same principle. This approach allows for better detection of third-party libraries within the application's non-main modules in subsequent analyses. Common module decoupling methods include using package structures, isomorphic graphs, program dependency graphs, class dependency graphs, and other information to identify independent modules. Some studies place emphasis on feature selection in the data processing phase. By selecting more representative and obfuscation-resistant features, it becomes possible to more accurately identify the third-party libraries used in Android applications during the library matching phase. Common features include third-party library names, class names, function names, character names, variable names, function centroids, package structures, class signatures, function signatures, control flow graphs, API call graphs, class dependency graphs, etc. Most research methods combine multiple features to form the final code features, allowing for a more comprehensive representation of the code information. Additionally, Android applications are often subjected to obfuscation, resulting in changes to identifiers and package structures. Using a combination of multiple features also enhances obfuscation resistance.

Early whitelist-based methods for detecting third-party libraries in Android applications\cite{zhou2012detecting,aafer2013droidapiminer,crussell2012attack} are a simple and direct approach that relies on a known whitelist of third-party libraries. This method matches the package names in the application with the names of TPL packages listed in the whitelist. If the names are the same or similar, it is inferred that the target application uses that TPL. However, due to the vastness of the Android ecosystem, it is challenging to establish a comprehensive whitelist of TPLs. Additionally, Android applications often employ obfuscation techniques, such as renaming packages, classes, and methods. Detecting third-party libraries in software solely based on package names is not accurate and may lead to a significant number of false negatives.

Based on the shortcomings of early research methods, researchers have proposed source code-based detection methods. Soh et al.\cite{soh2016libsift} introduced LibSift, an Android application third-party library detection tool based on program dependency graphs. LibSift supports extracting program dependency graph features to counter common code obfuscation issues and can detect less popular third-party libraries. Specifically, LibSift primarily performs module decoupling at the package level, using program dependency graphs to decouple Android applications into separate modules. It then identifies whether a module is a main module based on the rule that third-party libraries are designed to work independently and should not depend on other third-party libraries to run, meaning non-main modules should not depend on other non-main modules. By leveraging the number of modules a module depends on, LibSift identifies whether it is a main module. Using PDG, LibSift can identify the dependency relationships of each module on other modules and calculate the total number of modules it depends on. However, it can only identify main modules and non-main modules in Android applications to determine candidates for third-party libraries and cannot identify specific third-party libraries.

LibSift can only identify whether modules in Android applications are third-party libraries and cannot recognize the names of third-party libraries. Although early whitelist-based methods can identify the names of third-party libraries used in Android applications, their accuracy is not high. Zhang et al.\cite{zhang2018detecting} proposed LibPecker, an Android library detection tool based on class dependency relationships and signature matching, which generates specific signatures for each class. In the library matching stage, adaptive similarity thresholds and weighted class similarity scores are introduced to address issues such as partial imports, modifications, and library updates when computing similarities. Han et al.\cite{han2018identify} extracted method-level features to build a hash tree. Specifically, Androguard was used to construct a control flow graph (CFG) for each method, from which method-level features including opcodes and Android type tags were extracted. These features were then hashed to obtain the hash value for each method. Subsequently, the hash values of all methods within the same class were sorted in ascending order to obtain the hash value for the class. Similarly, the hash values for packages and applications were obtained using the same method, and only the hash tree containing the application hash value, package hash value, and class hash value was stored. Finally, a similarity comparison method was employed to identify third-party libraries within the application. Previous methods had certain limitations, especially in cases where similar third-party libraries or multiple versions of third-party libraries existed, leading to high time complexity and low accuracy. To overcome this issue, Xu et al.\cite{xu2020libroad} proposed LibRoad, a fast, online, and accurate TPL detection tool that combines package matching strategies based on package names and signatures. The package matching strategy based on package names is used to match unobfuscated packages, while the combination of package matching strategies based on signatures is used to match obfuscated packages, resulting in lower time complexity. Additionally, LibRoad adopts package filtering mechanisms, online TPL detection, and local TPL discovery to identify TPLs with low false positives and false negatives.

The methods mentioned above cannot identify the versions of third-party libraries used in Android applications. Backes et al.\cite{backes2016reliable} proposed the first tool, called LibScout, capable of identifying the versions of third-party libraries used in applications. It supports resistance against common obfuscation techniques such as identifier renaming and API hiding. LibScout generates a package tree through Class Hierarchy Analysis (CHA) and applies hierarchical fuzzy hashing from top to bottom at the package, class, and method levels in the package tree. It then matches third-party libraries using the hash values. However, its code feature granularity is too coarse, and it only contains syntactic information, which can lead to false negatives. Titze et al.\cite{titze2017ordol} introduced Ordol, a third-party library detection tool built upon plagiarism detection methods, to detect specific library versions in applications in an obfuscation-resistant manner. Ordol combines control flow graphs and K-grams to extract code features and construct third-party library candidates. This method uses method-level and class-level code features to iteratively match with a third-party library database to identify the third-party library and its version. However, Ordol's dataset only includes 40 third-party libraries and 1000 Android applications, limiting its universality. Zhang et al.\cite{zhang2019libid} proposed LibID, an obfuscation-resistant, fully automatic tool for detecting Android libraries. It identifies candidate third-party library versions in applications through static analysis of application binary files and a third-party library database. LibID combines class access flags, superclass names, class interfaces, method descriptors, and basic block signatures to represent class signatures. It determines the third-party library versions used in applications through three steps: class matching, dependency matching, and library matching. However, the features it selects are not robust against more advanced obfuscation techniques. Xian et al.\cite{zhan2021atvhunter} proposed a two-stage detection method called ATVHunter to detect third-party libraries and their versions in Android applications. It extracts class control flow graphs as coarse granularity and opcode in each basic block as fine granularity for library matching, which exhibits some resistance to obfuscation. Additionally, ATVHunter constructs a third-party library vulnerability dataset to identify 1-day vulnerabilities based on the detected third-party library database versions. Current detection tools for third-party libraries in Android applications suffer from poor scalability, low accuracy for fast tools, and high detection time costs for accurate tools. Huang et al.\cite{huang2022scalably} proposed an extensible two-stage detection method called Libloom. It abstracts third-party library detection as a set problem, extracting signature sets from packages and classes and encoding them into two types of Bloom filters. Libloom uses package-level Bloom filters to identify candidate third-party libraries and class-level filters to identify accurate third-party libraries. It also employs an entropy-based method to identify repackaged or obfuscated packages, where packages with maximum entropy are likely to be repackaged or obfuscated. Libloom can also identify the versions of third-party libraries used.

The methods mentioned above for detecting the versions of third-party libraries are all based on Android bytecode or source code. Tang et al.\cite{tang2022libdb} designed a binary-level third-party library detection tool called LibDB, which extracts string constants, exported function names, and function control flow graphs as basic features. Further, it embeds function features into a neural network to obtain function vectors. In the third-party library matching stage, LibDB first uses basic features and function vectors to obtain candidate third-party libraries. Then, it matches each candidate library with the target software using a Function Call Graph (FCG) filter to determine the used third-party library names. Finally, it determines the third-party library versions based on the similarity score.

The aforementioned methods only focus on Java libraries in Android applications. Some researchers also pay attention to native libraries in addition to Java libraries. Duan et al.\cite{duan2017identifying} proposed OSSPolice, which extracts features from binary files of applications for third-party library detection. It can detect both Java and native libraries in applications. Based on the detected results, OSSPolice can identify open-source software license conflicts and 1-day vulnerabilities in applications. The researchers of OSSPolice consider third-party libraries as modular and organized. They use the package structure as a feature for module decoupling to construct candidate third-party libraries. They extract multiple features such as string constants, control flow graphs, and fuzzy signatures as code features. They also introduce a new hierarchical indexing scheme for third-party library detection. However, OSSPolice cannot identify third-party libraries in applications if the package structure has been modified, and it has weak resistance to obfuscation. Most existing methods rely on syntactic features, so when these features are obfuscated, the detection accuracy is compromised. Additionally, existing methods are mostly based on the assumption that third-party libraries are fully imported into the software and cannot detect cases of partial import. Addressing these issues, Yang et al.\cite{yang2022modx}] proposed ModX, a framework that uses novel program modularization techniques to decompose programs into modules based on fine-grained functions. ModX performs similarity comparison by extracting syntactic features, graph features, and function features of program modules and third-party library modules to obtain a list of used third-party libraries.

In addition to detection tools for general Android applications, there are also detection tools for specific types of Android applications, such as those targeting existing commercial software, which refers to software that has already been developed and is available for purchase, as opposed to custom development. Feng et al.\cite{yuan2019b2sfinder} proposed a novel binary-to-source matching tool called B2SFinder for detecting third-party libraries in commercial-off-the-shelf (COTS) software, introducing the concept of reuse types to improve detection accuracy. Specifically, B2SFinder uses identifiers, constants, and control flow graphs as code features and employs a weighted feature matching algorithm. This algorithm combines three different code feature matching methods and two importance weighting methods to calculate the weight of a given code feature instance in COTS software based on its specificity and frequency in software applications. Finally, B2SFinder identifies different types of third-party library reuse based on matching scores from open-source software projects and code structure. However, the matching scores used by B2SFinder are based solely on similarity, and the selected features cannot effectively distinguish between versions of third-party libraries, thus failing to accurately identify third-party library versions. To address this issue, Ban et al.\cite{ban2021b2smatcher} proposed a two-stage commercial-off-the-shelf software third-party library version identification tool called B2SMatcher. This tool uses program-level code features as coarse-grained features and function-level code features as fine-grained features, and innovatively employs machine learning methods to obtain code features of source files involved in compilation. Additionally, it uses function abstraction and normalization methods to eliminate the comparison cost of redundant functions across versions.

Since the target software and third-party libraries may come from different platforms, applications and libraries on different platforms have different compilation and packaging formats, and the above tools are limited to the Android platform. Tang et al.\cite{tang2020libdx} proposed a platform-independent automated tool called LibDX, which can overcome the compilation diversity between binary files. LibDX extracts static constants and fuzzy filenames from the read-only DATA segment of binary files as code features to overcome the compilation and packaging formats on different platforms. It also considers the context of all strings in the target software, grouping them into multiple logical feature blocks, which are used to filter out library files and detect duplicate features in the target that do not have similar contexts, further reducing false positives caused by code features.

The above methods struggle to identify specific third-party library code fragments imported into the target software. Moreover, most of these methods involve string-based feature extraction, but strings often repeat across different third-party libraries, and there are numerous similar but non-isomorphic functions. These factors contribute to the unsatisfactory performance of most methods. Li et al.\cite{li2023libam} proposed a region matching framework called LibAM to detect third-party libraries and reused portions within the binary files of target software. LibAM connects isolated functions to function regions on the function call graph (FCG) and detects third-party libraries by comparing the similarity of these function regions, thereby significantly mitigating the impact of different optimization options and architectures.

Based on the above research, it can be observed that current research solutions mostly can only resist low-level code obfuscation techniques, while they still struggle with more advanced code obfuscation techniques. Additionally, the rapid growth and updates of third-party libraries make it challenging to maintain a comprehensive library database. Moreover, the methods mentioned above can only detect third-party libraries included in the database, leaving closed-source libraries and those not collected in the database undetected, posing a potential security risk to target software. Currently, only a small portion of research can utilize detection results to discover 1-day vulnerabilities and license conflicts, indicating that further research is needed in this direction. Lastly, since most third-party libraries are imported after modification, if the modified parts contain malicious code, it may lead to false positives in identifying 1-day vulnerabilities based on library version recognition.

2) Non-Android application software third-party library detection techniques

Research on third-party library detection techniques for non-Android application software is limited, primarily focusing on C/C++ software and iOS applications. These detection techniques often revolve around code segmentation to address the challenge of nested third-party libraries, which can lead to false positives. For instance, if a target software utilizes third-party library "a," but mistakenly identifies another library "b" due to nesting, it results in a false positive.

Lopes et al.\cite{lopes2017dejavu} proposed DejaVu, a tool based on code clone detection techniques aiming to analyze software dependencies among GitHub repositories by detecting project-level clones. However, this method solely relies on similarity comparison and thresholding, which is inaccurate, doesn't support large-scale software detection, and fails to identify nested third-party libraries. Woo et al.\cite{woo2021centris} introduced CENTRIS, an extensible tool capable of recognizing nested third-party libraries. CENTRIS employs code segmentation techniques, utilizing Locality-Sensitive Hashing (LSH) matching and thresholding to determine the third-party libraries used by the target software. Code segmentation ensures the removal of borrowed code (parts of nested third-party libraries), retaining only the application-specific code, thereby reducing false positives.

There is limited research on third-party library reuse in iOS applications. However, the potential security impact of third-party libraries in iOS applications cannot be overlooked. Guo et al.\cite{guo2022ilibscope} focused on detecting third-party libraries in iOS applications and proposed iLibScope, a configuration-file-based similarity comparison method. iLibScope constructs configuration files for each class and method of the target software and each version of the third-party library. It utilizes class-level configuration files for coarse-grained third-party library detection and method-level configuration files for fine-grained library version detection. iLibScope can also be applied to identify vulnerable third-party libraries in iOS applications, successfully identifying 405 vulnerable library uses from 4249 applications.

In summary, research on third-party library detection in C/C++ software and iOS applications is limited, yet the security risks posed by these libraries in such software categories are significant. Future research should pay more attention to detecting third-party libraries in these software categories. Challenges in researching C/C++ software detection may include the absence of a unified package manager, the vast and continually evolving landscape of C/C++ third-party libraries distributed across various platforms, and the difficulty in building and maintaining a comprehensive third-party library database. Existing databases of C/C++ third-party libraries are incomplete and lack real-time maintenance, warranting further research in this area.

(2) Unsupervised learning-based third-party library detection techniques

Unsupervised learning-based third-party library detection techniques, as opposed to those relying on known libraries, do not require dependency on third-party library databases. They can detect third-party libraries that may not be present in such databases, such as closed-source libraries, which is of significant importance for software security. If a third-party library carries malicious code, all applications integrating this library may be considered at risk. Known-library detection techniques, on the other hand, can only detect a limited set of third-party libraries, whereas unsupervised methods have the potential to detect additional ones.

Ma et al.\cite{ma2016libradar} introduced LibRadar in 2016, a tool based on API for third-party library detection. Stable APIs are extracted as code features and converted into hash values. Multiple clustering levels are applied to a large number of applications to identify potential third-party libraries. The same treatment is applied to the target software to generate hash values, which are then compared with those of potential libraries to detect third-party libraries. Although LibRadar can identify third-party libraries not included in known databases, it requires users to manually add library information such as names, categories, and official websites, which is inconvenient for users seeking information on third-party libraries used by target software.

Previous studies have mostly focused on code features without considering dependencies between packages. Li et al.\cite{li2017libd}] proposed LibD, which uses a graph-based method to identify third-party libraries in applications. LibD extracts information from multiple levels (packages, methods, and classes) and their dependencies to generate isomorphic graphs. These graphs are used for module decoupling, constructing candidate third-party libraries. Method-level control flow graphs (CFGs) are extracted for each potential library instance, generating method-level hash values from the CFGs. These hash values are combined to form class-level features, and clustering is performed based on these features. Candidate libraries with the same feature values are grouped together. If the number of candidates in a group equals or exceeds a predefined threshold, the group is considered a third-party library.

Liu et al.\cite{liu2022demystifying} proposed a third-party library clustering algorithm based on API call graphs, constructing potential third-party libraries from API call graphs in Android applications. GAT and CNN are used as similarity calculation models to compute the similarity between potential third-party libraries, followed by clustering using the DBSCAN algorithm.

Zhang et al.\cite{zhang2020empirical} introduced an automated tool called LibExtractor to identify third-party libraries in Android applications. It constructs candidate libraries based on dependencies between classes and employs a special four-round calculation algorithm to generate class and package feature values. These feature values are used to cluster candidate libraries, using a method similar to LibD. In addition to extracting third-party libraries, LibExtractor can identify potential malicious libraries. Researchers defined eight types of harmful behaviors that may exist in malicious libraries, allowing LibExtractor to detect whether the identified third-party libraries are malicious.

Zhang et al.\cite{zhang2021understanding} proposed LibHawkeye in 2021, a tool for detecting third-party libraries in large-scale Android applications. It uses a cluster-based technique to extract four types of dependencies (including package inclusion, inheritance, package homogeneity, and function calls) from application software to construct dependency graphs. Candidate libraries are then constructed based on these graphs, and method, class, and package features are generated sequentially and used for clustering to identify third-party libraries. LibHawkeye relies on internal dependency relationships within application software to construct candidate third-party libraries. However, if multiple subgraphs or different libraries are aggregated in one library, some false positives may occur.

Cui et al.\cite{cui2022libhunter} introduced LibHunter in 2022, which extracts runtime traffic data from application software and records corresponding API calls. Features are extracted from these two types of data and fed into a clustering algorithm to identify third-party libraries. Results showed that LibHunter performs better in detecting third-party libraries with network behaviors compared to existing methods. However, LibHunter's ability to detect certain categories of third-party libraries is limited. If a third-party library does not exhibit network behavior, there may be some false negatives.

In summary, unsupervised learning-based third-party library detection techniques offer significant contributions to software security by not requiring prior knowledge or maintenance of third-party library databases. However, existing unsupervised methods cannot provide specific information about the identified third-party libraries, such as their names or versions. Thus, it remains challenging to determine whether third-party libraries contain 1-day vulnerabilities, leaving ample room for further research in this area. Additionally, both known-library detection and unsupervised detection techniques can only detect vulnerabilities in known third-party libraries, lacking the ability to discover new vulnerabilities.

\section{Application of Third-Party Library Hardening Defense}

In today's software development, third-party libraries play a crucial role, providing developers with rich functionality and convenient solutions. However, as the widespread use of third-party libraries continues, associated security risks and potential vulnerabilities are increasingly evident. Malicious attackers can exploit vulnerabilities in third-party libraries to infiltrate systems, execute unauthorized operations, or steal sensitive information, posing a serious threat to software security. To address this growing security challenge, hardening defense of third-party libraries in software becomes essential. Hardening defense aims to strengthen the security of third-party libraries, mitigate potential risks, and enhance the overall security and stability of software. By taking appropriate hardening measures, developers can effectively prevent and mitigate security threats that may arise from third-party libraries, protect user data and privacy, and avoid system attacks and losses.

In this section, we will explore how to implement hardening defense for third-party libraries in software. We will consider methods of hardening defense from different levels and perspectives, including third-party library vulnerability patching, dependency management, license management, and other aspects. Additionally, we will study existing empirical research literature to understand hardening practices and effectiveness evaluations in different programming languages and ecosystems, aiming to provide developers with valuable resources and guidance to help them build more secure and reliable software. Through an in-depth exploration of hardening defense for third-party libraries in software, we hope to provide practical and valuable insights for researchers and practitioners in the software development and security fields, promoting the healthy development of software ecosystems together.

Timely patching of third-party library vulnerabilities is an essential step in hardening defense for third-party libraries in software. Security patches for vulnerabilities refer to secure updates released after fixing known vulnerabilities in third-party libraries. These patches effectively eliminate potential security risks and help software developers keep systems up-to-date and secure. Tan et al.\cite{tan2021locating} proposed PatchScout, a tool that utilizes vulnerability-commit relevance ranking to locate security patches for open-source software vulnerabilities. It ranks code commits based on multiple relevant features between a given vulnerability and code commits, helping users find security patches from a large number of code commits. Zhou et al.\cite{zhou2021finding} studied automatic mining of silent vulnerability patches and proposed VulFixMiner, a Transformer-based tool capable of identifying silent vulnerability fixes from code changes. Silent vulnerability fixes refer to vulnerabilities that have been fixed in code repositories before being publicly disclosed, posing security risks to users of open-source software, as malicious attackers may infer vulnerabilities from code changes and exploit them. VulFixMiner uses CodeBERT as a pre-trained language model to represent semantic changes at the file level, and then uses a neural network classifier to classify commit-level code changes. VulFixMiner was evaluated on Java and Python projects, showing superior performance in AUC and two performance metrics considering inspection costs compared to various existing baseline methods. VulFixMiner can also discover unreported vulnerability patches, i.e., vulnerability patches not recorded in public vulnerability databases. Nguyen et al.\cite{nguyen2022vulcurator} proposed VulCurator, a tool that uses deep learning and multiple information sources (including commit information, code changes, and issue reports) to detect vulnerability fix commits. Specifically, VulCurator first retrieves vulnerability information based on CVE numbers from public vulnerability databases such as NVD, Debian, and Red Hat, then extracts URL references from this information to construct a reference network representing resource references in vulnerability reporting, discussion, and resolution processes. Next, it selects commit nodes with high confidence and connectivity from the reference network as vulnerability fix commits. Finally, for each selected commit node, it expands the vulnerability fix commit collection by searching for relevant commits in different branches of its repository to address the possibility of multiple fix commits for one vulnerability. The paper evaluated VulCurator's performance through experiments and compared it with existing tools, showing significant advantages in accuracy and completeness. Xu et al. \cite{xu2022tracking} proposed an automated approach called TRACER to find patches for open-source software vulnerabilities from multiple sources. TRACER constructs a reference network to simulate resource references in vulnerability reporting, discussion, and resolution processes, then selects the most likely patch nodes based on the credibility and connectivity of nodes in the reference network, and finally expands the selected patch nodes by searching for relevant commits in different branches of the same repository to establish a one-to-many mapping between vulnerabilities and patches. Xie et al.\cite{xie2023precise} proposed a system called PHunter, which can accurately detect the presence of vulnerability fixes in obfuscated applications. PHunter's workflow includes three steps: candidate method localization, path extraction and pruning, and path summarization and comparison. PHunter uses coarse-grained and fine-grained semantic features to compare code before and after obfuscation, thereby resisting interference from code obfuscation. Additionally, PHunter can help eliminate false positives generated by existing third-party library detection tools, improving the efficiency of security analysis.

Locating and fixing vulnerabilities in third-party libraries is an important step in ensuring software security and stability. With the continuous development of automated vulnerability mining methods such as PatchScout, VulFixMiner, VulCurator, TRACER, and PHunter, software developers and security experts can more efficiently find security patches and promptly eliminate potential security risks. These methods leverage techniques such as deep learning, semantic representation, and reference network construction to improve the accuracy of vulnerability fixes and the ability to discover unreported vulnerability patches. However, ongoing research and improvement are still needed to address challenges such as the efficiency of locating vulnerabilities in large-scale software libraries and dealing with cases where one vulnerability may have multiple fix commits. By continuously improving the level of vulnerability fixing techniques and tools, we can better protect software systems from security threats and enhance the overall security of the software ecosystem.

Dependency management is a crucial measure in software development to ensure system security. By effectively managing and controlling third-party libraries and components, developers can reduce the risks of version conflicts and security vulnerabilities, while improving code maintainability and scalability. In practice, researchers have proposed a series of methods and tools to optimize the dependency management process, ensuring the security and stability of software systems. Vasilakis introduced an automated application isolation technique called BREAKAPP, based on module boundaries, aiming to enhance the security and reliability of software systems. BREAKAPP decomposes applications into protected isolation units leveraging the presence of third-party modules and enforces security policies. It creates isolation units at runtime and maintains the semantics of the original application through remote procedure calls (RPC). It also allows users to adjust the type of isolation units, communication methods, and functionality restrictions using optional policy expressions, thereby balancing security and performance trade-offs according to different needs and scenarios. Nguyen et al. designed Up2Dep, a tool to assist Android developers in keeping project dependencies up-to-date and avoiding the use of third-party libraries with security issues. Up2Dep is an Android Studio extension that detects and updates third-party libraries in Android projects, enhancing project security and compatibility. It utilizes an offline database to store API and encryption information of third-party libraries, providing different update suggestions and quick-fix solutions, while collecting feedback from developers. Vasilakis et al. proposed a new method to eliminate vulnerabilities in software components using Active Learning and Reconstruction (ALR) technology. ALR explores the behavior of components in a controlled environment, learns their observable functional models from the client side, and then reconstructs a new version of the component using that model. They also introduced a Domain Specific Language (DSL) for capturing string computations and designed an inference algorithm based on the DSL. They implemented an ALR system called Harp, which automatically rebuilds vulnerable string libraries, including those written in JavaScript and C/C++. Experimental results demonstrate that Harp can rapidly complete the rebuilding process in most cases while maintaining compatibility and performance with the original library. They also showed that Harp can eliminate several large-scale software supply chain attacks targeting string libraries.

The methods and tools mentioned above are extremely helpful in enhancing the security and reliability of software systems, providing crucial measures for effectively managing and controlling dependencies, especially in today's development environments where third-party libraries and components are widely used. Through these measures, software projects can be better protected from potential security threats.

License management is also an essential component of fortification defense. Third-party libraries often use different licenses, which may lead to conflicts or legal risks. Developers need to understand the licensing situation of third-party libraries and ensure that their licenses are compatible with those of the software project. Researchers have studied open-source software licenses. Mathur et al. focused on open-source software license conflicts, with the main purpose of detecting license conflicts and violations through code reuse. Xu et al. proposed a tool called LiDetector for detecting license incompatibility issues in open-source software, which can automatically understand license texts and infer rights and obligations to detect license incompatibility in open-source software.

In summary, the effective implementation of license management, dependency management, and vulnerability localization and repair will be key steps in ensuring the security of software systems. Furthermore, continuous advancement of technology and tools, along with strengthening developers' security awareness and knowledge training, will provide valuable support and assurance for the security of the software ecosystem. Only by addressing security issues such as dependency management and vulnerability repair comprehensively can software systems maintain a stable and trustworthy state in the constantly evolving technological environment.

\section{Analysis and Discussion}

\subsection{Summary and comparison}

In-depth research and analysis of the use of third-party libraries in software provide practical and valuable insights for developers, helping them build more secure and reliable software. By strengthening the management and maintenance of third-party libraries, developers can reduce potential risks associated with these libraries and ensure the overall security and reliability of software systems. Research into the ecosystems of third-party libraries for different programming languages reveals information about usage patterns, quality assessment, vulnerability propagation, and other aspects. These research findings offer beneficial resources and guidance for developers and researchers, enabling them to better understand and evaluate the characteristics and potential issues of third-party library ecosystems, thereby enhancing the overall security of software systems.

Regarding the detection techniques for third-party libraries in software, they offer better discovery of the libraries used in software and improved library management. Evaluation methods for these detection techniques mainly fall into two categories: one involves using common evaluation metrics such as accuracy, recall, and F1 score, while the other applies the tools to datasets constructed specifically for assessing whether they can detect third-party libraries in target software. However, due to the inconsistency in the primary contributions of various studies and the diversity in evaluation methods, with experiments often not based on a unified dataset, comparative experimental results cannot definitively determine the superiority or inferiority of models. Therefore, this paper takes a technical characteristic-oriented approach to compare and analyze various research efforts.

For each type of third-party library detection technique, this paper selects three representative research studies and summarizes them based on their technical characteristics. Table 1 presents these research studies, where each row represents one research work. The third column indicates the analyzed code form, including source code and binary code. The fourth column represents the analyzed programming semantics, including Java, C/C++, and Objective-C. The fifth column indicates the granularity of selection, including function-level, method-level, class-level, package-level, etc. The sixth column represents the feature selection, i.e., the features extracted from the analyzed objects that can represent those objects. The seventh column represents the matching method, i.e., how the extracted features are used to identify third-party libraries in the target software. The eighth column indicates whether the method can resist obfuscation, i.e., whether it can withstand code obfuscation techniques. The last column represents the supported recognition of third-party library risks, mainly including vulnerability identification, license conflicts, etc.

\begin{table*}
  \centering
  \caption{TPL detection technology.\label{tab1}}
  \scriptsize
  \begin{tabular}{@{}llllll@{}}\toprule 
    Tool & Language & Code Form & Feature extraction & Anti-obfuscation & Supported Risk Recognition \\  \midrule
    AtvHunter & Java & IR & Control Flow Graph & $\checkmark$ & 1day vulnerability \\
    ModX & C/C++ & Binary & Call Graph, String, Constant, Function Hash & $\times$ & $\times$ \\
    Centris & C/C++ & Source code & Function Hash & $\times$ & 1day vulnerability\\
    iLibScope & Objective-C & Binary & Class Name, Call Graph & $\times$ & 1day vulnerability\\
    LibD & Java & IR & Homogeny Graph, Control Flow Graph & $\checkmark$ & $\times$ \\
    LibHawkeye & Java & IR & Package Dependency Graph & $\checkmark$ & $\times$ \\ \bottomrule
  \end{tabular}
\end{table*}

By examining the representative research works, we can summarize and analyze as follows:

(1) In terms of research objectives, the research goals of third-party library detection techniques in software can be broadly divided into two categories: one category focuses solely on detecting whether modules in the software are third-party libraries. These detection methods can only determine whether a module in the software is a third-party library and cannot obtain detailed information about the third-party library, such as comparing library names and versions. However, these detection methods can detect closed-source third-party libraries and less popular third-party libraries. The other category not only focuses on detecting third-party libraries but also obtains detailed information about them. The advantage is that it can check for 1-day vulnerabilities and license conflicts based on the detailed information of third-party libraries. However, this method requires building a database of third-party libraries, and because it only focuses on the third-party libraries in the database, the detection results may not be comprehensive. Additionally, maintaining the third-party library database requires significant effort.

(2) Regarding the analyzed objects, current research primarily focuses on analyzing code and metadata to obtain features that can represent third-party libraries, which are then used for library matching. From Table 1, it can be observed that analysis of source code predominates. In terms of programming languages, there is a greater proportion of analysis conducted on Java programming language. There is also analysis on C/C++ and Objective-C, although to a lesser extent. The smaller proportion of research directly analyzing binary code may be due to differences in compilation platforms resulting in different binary outputs, and binary code may obscure some features of third-party libraries, making it difficult for researchers to overcome these obstacles for third-party library detection. The smaller proportion of analysis on C/C++ and Objective-C may be attributed to the lack of unified package managers for these languages, making it challenging to collect a comprehensive database of third-party libraries.

(3) Regarding feature selection, current research primarily extracts features such as control flow graphs, API call graphs, class dependency graphs, program dependency graphs, basic block/function/class/method/package signatures, string constants, function names, etc. Among these features, class dependency graphs, control flow graphs, and API call graphs contain syntactic information, while string constants and function names contain semantic information. Program dependency graphs and basic block/function/class/method/package signatures contain both syntactic and semantic information. Features that contain both semantic and syntactic information have higher performance for third-party libraries subjected to code obfuscation, enabling them to resist code obfuscation techniques.

(4) Regarding supported recognition of defects, the majority of existing research is limited to detecting third-party libraries in software, while a minority of research can identify risks present in third-party libraries. OSSPlice, for example, can identify open-source software license conflicts and 1-day vulnerabilities based on the detected results. ATVHunter also constructs a dataset of third-party library vulnerabilities that can identify 1-day vulnerabilities based on detected versions of third-party databases.

Through summarizing and comparing various representative research works, it can be observed that research on third-party library detection techniques in software has become a hotspot in the field of software security. Researchers conduct in-depth analysis of third-party library metadata and code, construct appropriate feature forms, and then match or cluster the third-party libraries in the target software, achieving a series of breakthrough results, mainly reflected in:

(1) More granular feature extraction: from early package-level extraction to recent studies achieving granularity at the level of basic blocks.
(2) More effective feature extraction: from early single-feature extraction to current research extracting multiple feature combinations for detection; from early extraction of only semantic features to current research extracting features that include both semantic and syntactic information, thereby improving resistance to obfuscation and accuracy.
(3) More in-depth research: from early extraction of third-party libraries in software to the current more in-depth identification of risks posed by detected third-party libraries. From early detection of only Java libraries to now being able to detect C/C++ libraries and Objective-C libraries.

This series of new theories and technologies proposed in third-party library detection continually expands the capabilities of third-party library detection.

Strengthening the defense of third-party libraries in software can be achieved through implementing license management, dependency management, and locating and fixing vulnerabilities, all of which can enhance the security of software systems. Continuous advancement in technological development and security awareness training for developers are also crucial for safeguarding the security of the software ecosystem. Only by considering these factors comprehensively can software systems maintain stability and trustworthiness in the ever-evolving technological environment.

\subsection{Problems and prospects}

Based on the above research, this article provides a detailed analysis of the challenges faced by third-party library security research in current application software.

Third-party library detection technology faces multiple challenges. (1) With the vigorous development of the open source community and software market, the number and complexity of third-party libraries continue to increase, making the detection process more difficult and prone to omissions or false positives. In addition, detection methods that rely on known libraries also increase the maintenance cost of third-party library databases, which need to be updated in time to maintain effectiveness. (2) Most existing research can only deal with some code obfuscation techniques, and as malicious actors continue to develop new obfuscation techniques, detection difficulty further increases. The continuous evolution of this technology requires simultaneous updates of detection technology to maintain effective protection against new threats. (3) Diversified trends in software development. As developers use different programming languages and technologies to build applications, third-party library detection technology needs to support various platforms, which increases the development and maintenance costs of detection tools. In order to meet this challenge, universal cross-language and cross-platform detection models have become the future development direction.

Third-party library detection technology needs to continue to evolve to adapt to the rapidly changing software environment. Deep learning technology will become a key tool to identify new malicious behaviors and vulnerabilities through training models, improving detection accuracy and efficiency. At the same time, universal cross-language and cross-platform detection models will become a trend to help cope with emerging new technologies and programming languages and reduce development and maintenance costs. Integration with software development and continuous integration tools will become a trend, and automated and preventive detection can help reduce potential risks and detect and solve problems in advance. In addition, strengthening the detection and protection of privacy- and security-sensitive code is an important direction for future development to ensure the security of user data and applications and maintain user trust. In summary, third-party library detection technology will continue to play an important role in the continuous challenges and developments, providing strong support for the safe and stable development of the software ecosystem.

Third-party libraries in application software face multiple challenges. (1) A large number and complex dependencies. The number of third-party libraries widely used in software development is huge, and there may be complex dependencies between them. Ensuring these libraries are secure and hardened is a complex task, as each library may require different security methods and tools. (2) Insufficient reinforcement of defense capabilities. Existing hardening and defense technologies may not be able to fully cope with unknown vulnerabilities and risks. Attackers are constantly developing new attack methods and techniques, so defensive measures must continue to evolve and innovate to remain effective. (3) Insufficient supervision of third-party libraries. Security issues in third-party libraries may affect a large number of downstream software. However, policing and managing the security of third-party libraries is insufficient in many cases. Stricter third-party library audits and security reviews are key to reducing risks. (4) Open source software license conflicts. The use of open source software is very common in modern software development. However, licensing open source software can involve complex legal and compliance issues. Ensuring compliance with all open source licenses is critical to avoiding potential legal risks.

To solve the challenges faced by third-party libraries, researchers can take the following measures: (1) Develop intelligent security tools to automatically detect and repair vulnerabilities in third-party libraries and respond to unknown threats. (2) Strengthen cooperation and information sharing in the security community, jointly respond to new security threats, and promptly disseminate information about vulnerabilities and risks. (3) Regularly update and review third-party libraries and promptly fix known vulnerabilities and security issues. (4) Ensure compliance with open source software licenses and avoid license conflicts. (5) Improve developers’ security training and awareness so that they are familiar with common security vulnerabilities and best practices to write more secure code and use third-party libraries correctly. These comprehensive measures will help improve software security and stability and protect applications from potential attacks and vulnerabilities.

This section provides an in-depth analysis of the multiple challenges faced by security research on third-party libraries in current application software. Third-party library detection technology faces challenges brought by the large number and complex dependencies, the continuous evolution of malicious obfuscation techniques, and the diversified trend of software development. To deal with these problems, future development directions include the development of intelligent detection technology, universal cross-language and cross-platform detection models, as well as strengthening integration with software development tools and protecting privacy-sensitive code. These measures will help improve the security of third-party libraries and provide solid support for the healthy development of the software ecosystem.

\bibliographystyle{unsrt}

\begin{thebibliography}{10}

\bibitem{synopsys2023}
Synopsys.
\newblock Open source security and risk analysis, 2023.

\bibitem{ladisa2023sok}
Piergiorgio Ladisa, Henrik Plate, Matias Martinez, and Olivier Barais.
\newblock Sok: Taxonomy of attacks on open-source software supply chains.
\newblock In {\em 2023 IEEE Symposium on Security and Privacy (SP)}, pages 1509--1526. IEEE, 2023.

\bibitem{sonatype2021}
Sonatype.
\newblock Q3 2021 state of the software supply chain report.
\newblock \url{www.sonatype.com/resources/state-of-the-software-supply-chain-2021}, 2021.

\bibitem{google2022}
Google.
\newblock Understanding the impact of apache log4j vulnerability.
\newblock \url{https://security.googleblog.com/2021/12/understanding-impact-of-apache-log4j.html}, Feb 2022.

\bibitem{zhan2021research}
Xian Zhan, Tianming Liu, Lingling Fan, Li~Li, Sen Chen, Xiapu Luo, and Yang Liu.
\newblock Research on third-party libraries in android apps: A taxonomy and systematic literature review.
\newblock {\em IEEE Transactions on Software Engineering}, 48(10):4181--4213, 2021.

\bibitem{heartbleed}
{Heartbleed}.
\newblock The heartbleed bug.
\newblock \url{https://heartbleed.com/}.

\bibitem{wang2020empirical}
Ying Wang, Bihuan Chen, Kaifeng Huang, Bowen Shi, Congying Xu, Xin Peng, Yijian Wu, and Yang Liu.
\newblock An empirical study of usages, updates and risks of third-party libraries in java projects.
\newblock In {\em 2020 IEEE International Conference on Software Maintenance and Evolution (ICSME)}, pages 35--45. IEEE, 2020.

\bibitem{prana2021out}
Gede Artha~Azriadi Prana, Abhishek Sharma, Lwin~Khin Shar, Darius Foo, Andrew~E Santosa, Asankhaya Sharma, and David Lo.
\newblock Out of sight, out of mind? how vulnerable dependencies affect open-source projects.
\newblock {\em Empirical Software Engineering}, 26:1--34, 2021.

\bibitem{gkortzis2021software}
Antonios Gkortzis, Daniel Feitosa, and Diomidis Spinellis.
\newblock Software reuse cuts both ways: An empirical analysis of its relationship with security vulnerabilities.
\newblock {\em Journal of Systems and Software}, 172:110653, 2021.

\bibitem{zajdel2022open}
Stan Zajdel, Diego~Elias Costa, and Hafedh Mili.
\newblock Open source software: an approach to controlling usage and risk in application ecosystems.
\newblock In {\em Proceedings of the 26th ACM International Systems and Software Product Line Conference-Volume A}, pages 154--163, 2022.

\bibitem{cao2022towards}
Yulu Cao, Lin Chen, Wanwangying Ma, Yanhui Li, Yuming Zhou, and Linzhang Wang.
\newblock Towards better dependency management: A first look at dependency smells in python projects.
\newblock {\em IEEE Transactions on Software Engineering}, 2022.

\bibitem{bommarito2019empirical}
Ethan Bommarito and Michael Bommarito.
\newblock An empirical analysis of the python package index (pypi).
\newblock {\em arXiv preprint arXiv:1907.11073}, 2019.

\bibitem{ruohonen2021large}
Jukka Ruohonen, Kalle Hjerppe, and Kalle Rindell.
\newblock A large-scale security-oriented static analysis of python packages in pypi.
\newblock In {\em 2021 18th International Conference on Privacy, Security and Trust (PST)}, pages 1--10. IEEE, 2021.

\bibitem{liu2022demystifying}
Chengwei Liu, Sen Chen, Lingling Fan, Bihuan Chen, Yang Liu, and Xin Peng.
\newblock Demystifying the vulnerability propagation and its evolution via dependency trees in the npm ecosystem.
\newblock In {\em Proceedings of the 44th International Conference on Software Engineering}, pages 672--684, 2022.

\bibitem{zerouali2022impact}
Ahmed Zerouali, Tom Mens, Alexandre Decan, and Coen De~Roover.
\newblock On the impact of security vulnerabilities in the npm and rubygems dependency networks.
\newblock {\em Empirical Software Engineering}, 27(5):107, 2022.

\bibitem{mir2023effect}
Amir~M Mir, Mehdi Keshani, and Sebastian Proksch.
\newblock On the effect of transitivity and granularity on vulnerability propagation in the maven ecosystem.
\newblock In {\em 2023 IEEE International Conference on Software Analysis, Evolution and Reengineering (SANER)}, pages 201--211. IEEE, 2023.

\bibitem{wu2023understanding}
Yulun Wu, Zeliang Yu, Ming Wen, Qiang Li, Deqing Zou, and Hai Jin.
\newblock Understanding the threats of upstream vulnerabilities to downstream projects in the maven ecosystem.
\newblock In {\em 2023 IEEE/ACM 45th International Conference on Software Engineering (ICSE)}, pages 1046--1058. IEEE, 2023.

\bibitem{tang2022towards}
Wei Tang, Zhengzi Xu, Chengwei Liu, Jiahui Wu, Shouguo Yang, Yi~Li, Ping Luo, and Yang Liu.
\newblock Towards understanding third-party library dependency in c/c++ ecosystem.
\newblock In {\em Proceedings of the 37th IEEE/ACM International Conference on Automated Software Engineering}, pages 1--12, 2022.

\bibitem{zhang2021study}
Haoxiang Zhang, Shaowei Wang, Heng Li, Tse-Hsun Chen, and Ahmed~E Hassan.
\newblock A study of c/c++ code weaknesses on stack overflow.
\newblock {\em IEEE Transactions on Software Engineering}, 48(7):2359--2375, 2021.

\bibitem{zhan2021atvhunter}
Xian Zhan, Lingling Fan, Sen Chen, Feng We, Tianming Liu, Xiapu Luo, and Yang Liu.
\newblock Atvhunter: Reliable version detection of third-party libraries for vulnerability identification in android applications.
\newblock In {\em 2021 IEEE/ACM 43rd International Conference on Software Engineering (ICSE)}, pages 1695--1707. IEEE, 2021.

\bibitem{zhang2019libid}
Jiexin Zhang, Alastair~R Beresford, and Stephan~A Kollmann.
\newblock Libid: reliable identification of obfuscated third-party android libraries.
\newblock In {\em Proceedings of the 28th ACM SIGSOFT International Symposium on Software Testing and Analysis}, pages 55--65, 2019.

\bibitem{soh2016libsift}
Charlie Soh, Hee Beng~Kuan Tan, Yauhen~Leanidavich Arnatovich, Annamalai Narayanan, and Lipo Wang.
\newblock Libsift: Automated detection of third-party libraries in android applications.
\newblock In {\em 2016 23rd Asia-Pacific Software Engineering Conference (APSEC)}, pages 41--48. IEEE, 2016.

\bibitem{zhou2012detecting}
Wu~Zhou, Yajin Zhou, Xuxian Jiang, and Peng Ning.
\newblock Detecting repackaged smartphone applications in third-party android marketplaces.
\newblock In {\em Proceedings of the second ACM conference on Data and Application Security and Privacy}, pages 317--326, 2012.

\bibitem{aafer2013droidapiminer}
Yousra Aafer, Wenliang Du, and Heng Yin.
\newblock Droidapiminer: Mining api-level features for robust malware detection in android.
\newblock In {\em Security and Privacy in Communication Networks: 9th International ICST Conference, SecureComm 2013, Sydney, NSW, Australia, September 25-28, 2013, Revised Selected Papers 9}, pages 86--103. Springer, 2013.

\bibitem{crussell2012attack}
Jonathan Crussell, Clint Gibler, and Hao Chen.
\newblock Attack of the clones: Detecting cloned applications on android markets.
\newblock In {\em Computer Security--ESORICS 2012: 17th European Symposium on Research in Computer Security, Pisa, Italy, September 10-12, 2012. Proceedings 17}, pages 37--54. Springer, 2012.

\bibitem{zhang2018detecting}
Yuan Zhang, Jiarun Dai, Xiaohan Zhang, Sirong Huang, Zhemin Yang, Min Yang, and Hao Chen.
\newblock Detecting third-party libraries in android applications with high precision and recall.
\newblock In {\em 2018 IEEE 25th International Conference on Software Analysis, Evolution and Reengineering (SANER)}, pages 141--152. IEEE, 2018.

\bibitem{han2018identify}
Hongmu Han, Ruixuan Li, and Junwei Tang.
\newblock Identify and inspect libraries in android applications.
\newblock {\em Wireless Personal Communications}, 103:491--503, 2018.

\bibitem{xu2020libroad}
Jian Xu and Qianting Yuan.
\newblock Libroad: Rapid, online, and accurate detection of tpls on android.
\newblock {\em IEEE Transactions on Mobile Computing}, 21(1):167--180, 2020.

\bibitem{backes2016reliable}
Michael Backes, Sven Bugiel, and Erik Derr.
\newblock Reliable third-party library detection in android and its security applications.
\newblock In {\em Proceedings of the 2016 ACM SIGSAC conference on computer and communications security}, pages 356--367, 2016.

\bibitem{titze2017ordol}
Dennis Titze, Michael Lux, and Julian Schuette.
\newblock Ordol: Obfuscation-resilient detection of libraries in android applications.
\newblock In {\em 2017 IEEE Trustcom/BigDataSE/ICESS}, pages 618--625. IEEE, 2017.

\bibitem{huang2022scalably}
Jianjun Huang, Bo~Xue, Jiasheng Jiang, Wei You, Bin Liang, Jingzheng Wu, and Yanjun Wu.
\newblock Scalably detecting third-party android libraries with two-stage bloom filtering.
\newblock {\em IEEE Transactions on Software Engineering}, 49(4):2272--2284, 2022.

\bibitem{tang2022libdb}
Wei Tang, Yanlin Wang, Hongyu Zhang, Shi Han, Ping Luo, and Dongmei Zhang.
\newblock Libdb: An effective and efficient framework for detecting third-party libraries in binaries.
\newblock In {\em Proceedings of the 19th International Conference on Mining Software Repositories}, pages 423--434, 2022.

\bibitem{duan2017identifying}
Ruian Duan, Ashish Bijlani, Meng Xu, Taesoo Kim, and Wenke Lee.
\newblock Identifying open-source license violation and 1-day security risk at large scale.
\newblock In {\em Proceedings of the 2017 ACM SIGSAC Conference on computer and communications security}, pages 2169--2185, 2017.

\bibitem{yang2022modx}
Can Yang, Zhengzi Xu, Hongxu Chen, Yang Liu, Xiaorui Gong, and Baoxu Liu.
\newblock Modx: binary level partially imported third-party library detection via program modularization and semantic matching.
\newblock In {\em Proceedings of the 44th International Conference on Software Engineering}, pages 1393--1405, 2022.

\bibitem{yuan2019b2sfinder}
Zimu Yuan, Muyue Feng, Feng Li, Gu~Ban, Yang Xiao, Shiyang Wang, Qian Tang, He~Su, Chendong Yu, Jiahuan Xu, et~al.
\newblock B2sfinder: Detecting open-source software reuse in cots software.
\newblock In {\em 2019 34th IEEE/ACM International Conference on Automated Software Engineering (ASE)}, pages 1038--1049. IEEE, 2019.

\bibitem{tang2020libdx}
Wei Tang, Ping Luo, Jialiang Fu, and Dan Zhang.
\newblock Libdx: A cross-platform and accurate system to detect third-party libraries in binary code.
\newblock In {\em 2020 IEEE 27th International Conference on Software Analysis, Evolution and Reengineering (SANER)}, pages 104--115. IEEE, 2020.

\bibitem{li2023libam}
Siyuan Li, Yongpan Wang, Chaopeng Dong, Shouguo Yang, Hong Li, Hao Sun, Zhe Lang, Zuxin Chen, Weijie Wang, Hongsong Zhu, et~al.
\newblock Libam: An area matching framework for detecting third-party libraries in binaries.
\newblock {\em ACM Transactions on Software Engineering and Methodology}, 33(2):1--35, 2023.

\bibitem{lopes2017dejavu}
Cristina~V Lopes, Petr Maj, Pedro Martins, Vaibhav Saini, Di~Yang, Jakub Zitny, Hitesh Sajnani, and Jan Vitek.
\newblock D{\'e}j{\`a}vu: a map of code duplicates on github.
\newblock {\em Proceedings of the ACM on Programming Languages}, 1(OOPSLA):1--28, 2017.

\bibitem{woo2021centris}
Seunghoon Woo, Sunghan Park, Seulbae Kim, Heejo Lee, and Hakjoo Oh.
\newblock Centris: A precise and scalable approach for identifying modified open-source software reuse.
\newblock In {\em 2021 IEEE/ACM 43rd International Conference on Software Engineering (ICSE)}, pages 860--872. IEEE, 2021.

\bibitem{guo2022ilibscope}
Jingyi Guo, Min Zheng, Yajin Zhou, Haoyu Wang, Lei Wu, Xiapu Luo, and Kui Ren.
\newblock ilibscope: Reliable third-party library detection for ios mobile apps.
\newblock {\em arXiv preprint arXiv:2207.01837}, 2022.

\bibitem{ma2016libradar}
Ziang Ma, Haoyu Wang, Yao Guo, and Xiangqun Chen.
\newblock Libradar: Fast and accurate detection of third-party libraries in android apps.
\newblock In {\em Proceedings of the 38th international conference on software engineering companion}, pages 653--656, 2016.

\bibitem{li2017libd}
Menghao Li, Wei Wang, Pei Wang, Shuai Wang, Dinghao Wu, Jian Liu, Rui Xue, and Wei Huo.
\newblock Libd: Scalable and precise third-party library detection in android markets.
\newblock In {\em 2017 IEEE/ACM 39th International Conference on Software Engineering (ICSE)}, pages 335--346. IEEE, 2017.

\bibitem{zhang2020empirical}
Zicheng Zhang, Wenrui Diao, Chengyu Hu, Shanqing Guo, Chaoshun Zuo, and Li~Li.
\newblock An empirical study of potentially malicious third-party libraries in android apps.
\newblock In {\em Proceedings of the 13th ACM Conference on Security and Privacy in Wireless and Mobile Networks}, pages 144--154, 2020.

\bibitem{zhang2021understanding}
Yanghua Zhang, Jice Wang, Hexiang Huang, Yuqing Zhang, and Peng Liu.
\newblock Understanding and conquering the difficulties in identifying third-party libraries from millions of android apps.
\newblock {\em IEEE Transactions on Big Data}, 8(6):1511--1523, 2021.

\bibitem{cui2022libhunter}
Huajun Cui, Guozhu Meng, Yueqi Li, Yuejun Li, Yan Zhang, Jiyan Sun, Dali Zhu, and Weiping Wang.
\newblock Libhunter: An unsupervised approach for third-party library detection without prior knowledge.
\newblock In {\em 2022 IEEE Symposium on Computers and Communications (ISCC)}, pages 1--7. IEEE, 2022.

\bibitem{tan2021locating}
Xin Tan, Yuan Zhang, Chenyuan Mi, Jiajun Cao, Kun Sun, Yifan Lin, and Min Yang.
\newblock Locating the security patches for disclosed oss vulnerabilities with vulnerability-commit correlation ranking.
\newblock In {\em Proceedings of the 2021 ACM SIGSAC Conference on Computer and Communications Security}, pages 3282--3299, 2021.

\bibitem{zhou2021finding}
Jiayuan Zhou, Michael Pacheco, Zhiyuan Wan, Xin Xia, David Lo, Yuan Wang, and Ahmed~E Hassan.
\newblock Finding a needle in a haystack: Automated mining of silent vulnerability fixes.
\newblock In {\em 2021 36th IEEE/ACM International Conference on Automated Software Engineering (ASE)}, pages 705--716. IEEE, 2021.

\bibitem{nguyen2022vulcurator}
Truong~Giang Nguyen, Thanh Le-Cong, Hong~Jin Kang, Xuan-Bach~D Le, and David Lo.
\newblock Vulcurator: a vulnerability-fixing commit detector.
\newblock In {\em Proceedings of the 30th ACM Joint European Software Engineering Conference and Symposium on the Foundations of Software Engineering}, pages 1726--1730, 2022.

\bibitem{xu2022tracking}
Congying Xu, Bihuan Chen, Chenhao Lu, Kaifeng Huang, Xin Peng, and Yang Liu.
\newblock Tracking patches for open source software vulnerabilities.
\newblock In {\em Proceedings of the 30th ACM Joint European Software Engineering Conference and Symposium on the Foundations of Software Engineering}, pages 860--871, 2022.

\bibitem{xie2023precise}
Zifan Xie, Ming Wen, Haoxiang Jia, Xiaochen Guo, Xiaotong Huang, Deqing Zou, and Hai Jin.
\newblock Precise and efficient patch presence test for android applications against code obfuscation.
\newblock In {\em Proceedings of the 32nd ACM SIGSOFT International Symposium on Software Testing and Analysis}, pages 347--359, 2023.

\bibitem{vasilakis2018breakapp}
Nikos Vasilakis, Ben Karel, Nick Roessler, Nathan Dautenhahn, Andr{\'e} DeHon, and Jonathan~M Smith.
\newblock Breakapp: Automated, flexible application compartmentalization.
\newblock In {\em NDSS}, 2018.

\bibitem{vasilakis2021supply}
Nikos Vasilakis, Achilles Benetopoulos, Shivam Handa, Alizee Schoen, Jiasi Shen, and Martin~C Rinard.
\newblock Supply-chain vulnerability elimination via active learning and regeneration.
\newblock In {\em Proceedings of the 2021 ACM SIGSAC Conference on Computer and Communications Security}, pages 1755--1770, 2021.

\bibitem{mathur2012empirical}
Arunesh Mathur, Harshal Choudhary, Priyank Vashist, William Thies, and Santhi Thilagam.
\newblock An empirical study of license violations in open source projects.
\newblock In {\em 2012 35th Annual IEEE Software Engineering Workshop}, pages 168--176. IEEE, 2012.

\bibitem{xu2023lidetector}
Sihan Xu, Ya~Gao, Lingling Fan, Zheli Liu, Yang Liu, and Hua Ji.
\newblock Lidetector: License incompatibility detection for open source software.
\newblock {\em ACM Transactions on Software Engineering and Methodology}, 32(1):1--28, 2023.

\bibitem{ban2021b2smatcher}
Gu~Ban, Lili Xu, Yang Xiao, Xinhua Li, Zimu Yuan, and Wei Huo.
\newblock B2smatcher: fine-grained version identification of open-source software in binary files.
\newblock {\em Cybersecurity}, 4:1--21, 2021.

\end{thebibliography}

\end{document}